\documentclass
[a4paper,amsfonts,amssymb,titlepage,showpacs,preprint,showkeys,nofootinbib,nobibnotes,prl,twocolumn,10pt,byrevtex,floats]{revtex4}%
\usepackage{amsfonts}
\usepackage{amsmath}
\usepackage{amssymb}
\usepackage{graphicx}%
\providecommand{\U}[1]{\protect\rule{.1in}{.1in}}

\begin{document}
\title{Edge modes in the fractional quantum Hall effect without extra\\edge fermions}
\author{Gabriel Di Lemos Santiago Lima}
\email{gabriellemos3@hotmail.com}
\author{Sebasti\~{a}o Alves Dias}
\email{tiao@cbpf.br}
\affiliation{Centro Brasileiro de Pesquisas F\'{\i}sicas, Rua Dr. Xavier Sigaud, 150,
22290-180, Rio de Janeiro, RJ, Brazil}
\altaffiliation{Corresponding author: S.A. Dias}


\begin{abstract}
We show that the Chern-Simons-Landau-Ginsburg theory that describes the
quantum Hall effect on a bounded sample is anomaly free and thus does not
require the addition of extra chiral fermions on the boundary to restore local
gauge invariance.

\end{abstract}
\maketitle

\section{Introduction}

From a field theoretical point of view, the fractional quantum Hall effect
(FQHE) is well understood since 1989, when it was shown that it could be
described by a Chern-Simons-Landau-Ginsburg (CSLG) effective theory
\cite{zhang}. Two years later, the effect of boundaries was analysed
\cite{wen} and a breakdown of gauge invariance was found, due to a bounded
Chern-Simons (CS) action. Noticing the fact that the microscopic theory
\textit{is} gauge invariant, several authors \cite{wen,maeda} concluded that
an anomaly appeared, that would have to be cancelled in order to recover gauge
invariance. This was done by the \textit{addition of extra degrees of
freedom}, in the form of chiral fermions circulating at the edge, whose well
known gauge anomaly restored gauge invariance of the complete theory. These
extra degrees of freedom were, {since} then, {usually} identified with chiral
edge modes, {which are} fundamental in the wave function approach to the FQHE
\cite{girvin}.

In this letter, we briefly review the steps conducting from the microscopic
Hamiltonian to the CSLG theory in the presence of boundaries. {We show that,
by carefully considering the influence of the boundary in the dynamics of the
CS field, one finds no such anomaly and the resulting theory is \textit{gauge
invariant}}. Thus, \textit{there is no need to introduce extra degrees of
freedom such as chiral edge fermions}. The edge modes appear naturally from
the dynamics of the CS field, which is determined by its coupling to the
Noether current of the CSLG theory. {Thus}, they have nothing to do with the
chiral edge fermions, introduced in previous approaches.

Edge states were found experimentally \cite{wang-goldman} and appear as an
essential ingredient in several recent works. We can quote some of them which
deal with various subjects such as edge states in graphene \cite{yao,hatsugai}%
, descriptions of chiral Luttinger liquids \cite{boyarsky,chang} and relations
between edge electrons and Berry's phase \cite{basu}. Accordingly, the
association of these modes with the {original} degrees of freedom of the CSLG
effective theory is very important both for present and future applications.

\section{FQHE on a bounded surface}

We briefly review the derivation of the CSLG model for the FQHE \cite{zhang},
modifying the procedure when necessary to take into account the finiteness of
the surface. We start from the microscopic Hamiltonian of a two-dimensional
system of polarized electrons interacting with an external electromagnetic
field\textbf{.}%

\begin{align}
&  H_{\mathrm{F}} =\frac{1}{2m}\sum\limits_{r}\left[  \mathbf{p}_{r}-\frac
{e}{c}\mathbf{A}(\mathbf{x}_{r})\right]  ^{2}+\sum\limits_{r}eA_{0}%
(\mathbf{x}_{r})\label{15}\\
&  +\sum\limits_{r<s} V(\mathbf{x}_{r}-\mathbf{x}_{s}) {+ \sum\limits_{r}%
V_{c}(\mathbf{x}_{r})}.\nonumber
\end{align}
{In (\ref{15}), $V_{c}(\mathbf{x}_{r})$ is an electrostatic potential which is
responsible for confining the electrons into the bounded region $\Gamma$. It
is essentially zero in the bulk of $\Gamma$ and very large as one approaches
the boundary.} As the electrons are polarized, the wave function is completely
antisymmetric. It is possible to map this fermionic problem into a bosonic
one. This can be done by means of an unitary transformation%

\begin{equation}
U=\exp\left(  -i\sum_{r<s}\frac{\theta}{\pi}\alpha_{rs}\right)  , \label{16}%
\end{equation}
were $\alpha_{rs}$ is the angle between $\mathbf{x}_{r}-\mathbf{x}_{s}$ with
an arbitrary direction that may be chosen as the $x$-axis and $\theta
=(2k+1)\pi$ with $k$ being an integer. Under this choice, one can easily
verify that an antisymmetric wave function $\psi$ is mapped into a symmetric
(bosonic) one $\phi\equiv U^{-1}\psi$. It is easy to check that%
\begin{equation}
U^{-1}\left(  \mathbf{p}_{r}-\frac{e}{c}\mathbf{A}(\mathbf{x}_{r})\right)  U
=\mathbf{p}_{r}-\frac{e}{c}\mathbf{A}(\mathbf{x}_{r})-\hbar\frac{\theta}{\pi
}\sum\limits_{r\neq s}\boldsymbol{\nabla}\alpha_{rs}. \label{17}%
\end{equation}
The gradient of the angle between the vector $\mathbf{x}_{r}-\mathbf{x}_{s}$
and the $x$-axis is given by%

\begin{equation}
\partial_{i}\alpha_{rs}\equiv\partial_{i}\alpha(\mathbf{x}_{r}\mathbf{-x}%
_{s})=-\varepsilon_{ij}\frac{x_{r}^{j}-x_{s}^{j}}{|\mathbf{x}_{r}%
-\mathbf{x}_{s}|^{2}}. \label{18}%
\end{equation}
Now, a statistical field is \textit{defined}%

\begin{equation}
\mathbf{a(x}_{r})\equiv\frac{\phi_{0}}{2\pi}\frac{\theta}{\pi}\sum_{s\neq
r}\boldsymbol{\nabla}\alpha_{rs}, \label{20}%
\end{equation}
where $\phi_{0}=hc/e$ is the quantum of flux. The bosonized Hamiltonian is
$H_{\mathrm{B}}=U^{-1}H_{\mathrm{F}}U$ or%

\begin{align}
&  H_{\mathrm{B}} =\frac{1}{2m}\sum\limits_{r}\left(  \mathbf{p}_{r}-\frac
{e}{c}[\mathbf{A}(\mathbf{x}_{r})+\mathbf{a}(\mathbf{x}_{r})]\right)
^{2}\label{21}\\
&  +\sum\limits_{r}eA_{0}(\mathbf{x}_{r})+\sum\limits_{r<s}V(\mathbf{x}%
_{r}-\mathbf{x}_{s}) {+\sum\limits_{r}V_{c}(\mathbf{x}_{r})}.\nonumber
\end{align}

The second quantized Hamiltonian is obtained through the introduction of a
bosonic field $\phi\left(  \mathbf{x}\right)  $ ($\mathbf{x}$ denotes $\left(
x_{1},x_{2}\right)  $) satisfying$\left[  \phi\left(  \mathbf{x}\right)
,{\phi}^{\dagger}{(\mathbf{y})}\right]  =\delta^{\left(  2\right)  }\left(
\mathbf{x}-\mathbf{y}\right)  , $ and generalizing $H_{\mathrm{B}}$ to the
(hermitian) \textit{matter Hamiltonian}
\begin{align}
&  H_{\mathrm{M}} =\int_{{\Gamma}}d^{2}x\left\{  \frac{\hbar^{2}}{2m}\left(
\mathbf{D}\phi\left(  \mathbf{x}\right)  \right)  ^{\dagger}\cdot
\mathbf{D}\phi\left(  \mathbf{x}\right)  \right\} \label{23}\\
&  +\int_{{\Gamma}}d^{2}x\left\{  eA_{0}\left(  \mathbf{x}\right)
\phi^{\dagger}\left(  \mathbf{x}\right)  \phi\left(  \mathbf{x}\right)
\right\} \nonumber\\
&  +\frac{1}{2}\int_{{\Gamma}}d^{2}x \, d^{2}y \, {\delta\rho}\left(
\mathbf{x}\right)  {V (\mathbf{x}-\mathbf{y})}{\delta\rho}\left(
\mathbf{y}\right)  .\nonumber
\end{align}
{Notice that we incorporated the effect of the confining potential $V_{c}$ by
restricting the integration domain over the region $\Gamma$. An alternative
approach would be to explicitly take into account the confining potential by
adding a Gaussian term $\int d^{2}x\left\{  V_{c}\left(  \mathbf{x}\right)
\phi^{\dagger}\left(  \mathbf{x}\right)  \phi\left(  \mathbf{x}\right)
\right\}  $ and not restricting integration. We will follow the first approach
to make it easier to compare our results with the literature.} In (\ref{23})
the covariant derivative was defined as%
\begin{equation}
D^{k}=\partial^{k}+\frac{ie}{\hbar c}\left(  A^{k}(\mathbf{x})+a^{k}%
(\mathbf{x})\right)  , \label{5}%
\end{equation}
$\rho\left(  \mathbf{x}\right)  ={\phi}^{\dagger}{(\mathbf{x})\phi
(\mathbf{x})}$ and $\delta\rho\left(  \mathbf{x}\right)  =\rho\left(
\mathbf{x}\right)  -\bar{\rho}$ (the average density $\bar{\rho}$ is included
here to avoid problems in the case of a long range potential \cite{zhang}).
Taking all operators in the Heisenberg picture, they become functions of time.
The action below generates the correct Heisenberg equations%
\begin{align}
&  S_{\mathrm{M}} =\int d^{3}x\left\{  \frac{i\hbar c}{2}\Theta\left(
\mathbf{x}\right)  \phi^{\dagger}\left(  x\right)  D_{0}\phi(x)\right\}
\label{5j}\\
&  +\int d^{3}x\left\{  - \frac{i\hbar c}{2}\Theta\left(  \mathbf{x}\right)
\left(  D_{0}\phi\left(  x\right)  \right)  ^{\dagger}\phi(x)\right\}
\nonumber\\
&  +\int d^{3}x\left\{  -\frac{\hbar^{2}}{2m}\Theta\left(  \mathbf{x}\right)
\left(  \mathbf{D}\phi\left(  x\right)  \right)  ^{\dagger}\cdot\mathbf{D}%
\phi(x)\right\} \nonumber\\
&  -\frac{1}{2}\int d^{3}x\, d^{3}y\, \Theta\left(  \mathbf{x}\right)
\Theta\left(  \mathbf{y}\right)  {\delta\rho}\left(  x\right)  {
V(\mathbf{x}-\mathbf{y} )}{\delta\rho\left(  y\right)  } ,\nonumber
\end{align}
with%
\begin{align}
&  D_{0} =\frac{1}{c}\partial_{t}+\frac{ie}{\hbar c}A_{0}\left(
\mathbf{x}\right) \label{5b}\\
&  \equiv\partial_{0}+\frac{ie}{\hbar c}A_{0}\left(  \mathbf{x}\right)
,\nonumber
\end{align}
and the integration being effectively over the surface of the sample, which is
obtained by the use of a step function $\Theta$ defined as%
\begin{equation}
\Theta\left(  \mathbf{x}\right)  =\left\{
\begin{array}
[c]{c}%
1, \; \; \mbox{if $\mathbf{x}\in {\Gamma}$},\\
0, \; \; \mbox{if $\mathbf{x}\notin {\Gamma}$}.
\end{array}
\right.  \label{5i}%
\end{equation}
Global phase invariance of the action (\ref{5j}) under the transformations
$\phi^{\prime}\left(  x\right)  =e^{i\alpha}\phi\left(  x\right)  $ implies
the continuity equation
\begin{equation}
\partial_{\mu}j_{\mathrm{M,}{\Gamma}}^{\mu}=\partial_{0}j_{\mathrm{M},{\Gamma
}}^{0}+\partial_{i}j_{\mathrm{M},{\Gamma}}^{i}\left(  x\right)  =0, \label{5k}%
\end{equation}
where the components of the matter current are given by%
\begin{align}
&  j_{\mathrm{M},{\Gamma}}^{0} =\Theta\left(  \mathbf{x}\right)  \phi
^{\dagger}\left(  x\right)  \phi\left(  x\right)  =\Theta\left(
\mathbf{x}\right)  \rho\left(  x\right)  ,\nonumber\\
&  \mathbf{j}_{\mathrm{M},{\Gamma}} = \Theta\left(  \mathbf{x}\right)
\mathbf{j}_{\mathrm{M}}\left(  x\right)  , \label{5g}%
\end{align}
with $\mathbf{j}_{\mathrm{M}}$ given by
\begin{equation}
\mathbf{j}_{\mathrm{M}}\left(  x\right)  = \frac{i\hbar}{2m}\left\{  {\phi
}^{\dagger}\left(  x\right)  \mathbf{D}{\phi}\left(  x\right)  -\left(
\mathbf{D}{\phi}\left(  x\right)  \right)  ^{\dagger}\phi\left(  x\right)
\right\}  .\nonumber
\end{equation}

The field $\mathbf{a}\left(  x\right)  $ is completely determined in terms of
the density operator $\rho\left(  x\right)  $, and is given by the second
quantized version of equation (\ref{20}),%
\begin{align}
&  a_{i}(x) =-\frac{\phi_{0}}{2\pi}\frac{\theta}{\pi}\varepsilon_{ij}%
\int_{{\Gamma}}d^{2}y\frac{x_{j}-y_{j}}{|\mathbf{x}-\mathbf{y}|^{2}}%
\rho\left(  y\right) \label{5h}\\
&  =-\frac{\phi_{0}}{2\pi}\frac{\theta}{\pi}\varepsilon_{ij}\int d^{2}%
y\frac{x_{j}-y_{j}}{|\mathbf{x}-\mathbf{y}|^{2}}\Theta\left(  \mathbf{y}%
\right)  \rho\left(  y\right)  .\nonumber
\end{align}
The field $\mathbf{a}\left(  x\right)  $ shown above can be seen as an
auxiliary field. It is the solution of the following pair of equations%
\begin{align}
&  \varepsilon_{ij}\partial_{i}a_{j}(x) =\phi_{0}\frac{\theta}{\pi}%
\Theta\left(  \mathbf{x}\right)  \rho(x),\label{5c}\\
&  \partial_{i}a_{i} = 0.\nonumber
\end{align}
Using the continuity equation (\ref{5g}) we can derive a third equation for
the field $a_{i}\left(  x\right)  $ involving a time derivative%
\begin{equation}
\varepsilon_{ij}\partial_{0}a_{j}(x)=-\phi_{0}\frac{\theta}{\pi}\Theta\left(
\mathbf{x}\right)  j_{\mathrm{M}}^{i}. \label{5d}%
\end{equation}
Equations (\ref{5c}) and (\ref{5d}) may be viewed as the equations of motion
of a new dynamical field, if we make the substitution $D_{0}=\partial
_{0}+\frac{ie}{\hbar c}A_{0}\left(  \mathbf{x}\right)  \rightarrow\partial
_{0}+\frac{ie}{\hbar c}\left(  A_{0}\left(  \mathbf{x}\right)  +a_{0}\left(
x\right)  \right)  $ (which means that now $S_{\mathrm{M}}=S_{\mathrm{M}%
}\left(  a_{0}\right)  $), and replace $S_{\mathrm{M}}$ by the action%
\begin{equation}
S=S_{\mathrm{M}}\left(  a_{0}\right)  +S_{\mathrm{CS}}, \label{action}%
\end{equation}
with%
\begin{equation}
S_{\mathrm{CS}}\equiv\frac{\sigma_{xy}}{2}\int d^{3}x \, \varepsilon^{\mu
\nu\rho}a_{\mu}\partial_{\nu}a_{\rho}, \label{5e}%
\end{equation}
where $S_{\mathrm{CS}}$ is known as the Chern-Simons (CS) action and we define
the Hall conductivity as%
\begin{equation}
\sigma_{xy}\equiv\frac{\pi}{\theta}\frac{1}{\phi_{0}}.
\end{equation}
The additional field $a_{0}\left(  x\right)  $ introduced in (\ref{5e}) and in
$S_{\mathrm{M}}$ can be eliminated by requiring the condition $a_{0}\left(
x\right)  =0$, what is legitimate in the context of a gauge field theory.
\textit{Thus, gauge invariance is crucial for the correct introduction of the
statistical CS field} $a_{\mu}\left(  x\right)  $. Without it, the equations
of motion obeyed by the field are not enough to eliminate this extra component
and the identification of the dynamics of $a_{\mu}\left(  x\right)  $ with
that of a CS field cannot be made.

The action $S$ is gauge invariant, because the CS part $S_{\mathrm{CS}}$
(equation \ref{5e}) is not being integrated over a finite surface. Then, under
a gauge transformation,%
\begin{align}
&  S_{\mathrm{CS}}[a_{\mu}+\partial_{\mu}\alpha]=S_{\mathrm{CS}}[a_{\mu
}]+\frac{\sigma_{xy}}{2}\int d^{3}x \, \varepsilon^{\mu\nu\rho}\left(
\partial_{\mu}\alpha\right)  \partial_{\nu}a_{\rho}\nonumber\\
&  =S_{\mathrm{CS}}[a_{\mu}]-\frac{\sigma_{xy}}{2}\int d^{3}x \, \partial
_{\mu}\left(  \varepsilon^{\mu\nu\rho}\alpha\partial_{\nu}a_{\rho}\right)
\nonumber\\
&  =S_{\mathrm{CS}}[a_{\mu}].
\end{align}
The restriction of the domain of integration to the area of the sample is the
origin of the destruction of gauge invariance \cite{wen,maeda}. This problem
is absent here. Thus, gauge invariance is a consequence of unbounded
integration in equation (\ref{5e}).

If one insisted to consider the CS action on a surface with a boundary,%
\begin{equation}
S_{\mathrm{CS}}^{b}\equiv\frac{\sigma_{xy}}{2}\int d^{3}x \, \Theta\left(
\mathbf{x}\right)  \varepsilon^{\mu\nu\rho}a_{\mu}\partial_{\nu}a_{\rho
},\nonumber
\end{equation}
the resulting equations of motion for the CS field would be (setting $a_{0}=0$
artificially, as this is only allowed if the theory is gauge invariant)%
\begin{align}
&  \varepsilon_{ij}\partial_{i}a_{j}(x) =\phi_{0}\frac{\theta}{\pi}\rho(x),\\
&  \varepsilon_{ij}\partial_{0}a_{j}(x) =-\phi_{0}\frac{\theta}{\pi}j^{i},
\end{align}
and these equations would result in the solution%
\begin{equation}
a_{i}(x)=-\frac{\phi_{0}}{2\pi}\frac{\theta}{\pi}\varepsilon_{ij}\int
d^{2}y\frac{x_{j}-y_{j}}{|\mathbf{x}-\mathbf{y}|^{2}}\rho\left(  y\right)
,\nonumber
\end{equation}
\textit{which does not contain the restriction of the integration to the
surface of the sample and, so, does not coincide with (\ref{5h})}. This would
result in a second quantized action which is not equivalent to the original
problem. Therefore another reason why the CS action \textit{must} be unbounded
is to provide the correct solution for the statistical field.

Although the Chern-Simons part of the action makes no reference to the
boundary of the sample, we have to remember that the CS field is minimally
coupled to the matter fields. This coupling involves the Noether current and
forces the equations of motion for the CS field to depend on the boundary, as
can be explicitly seen in (\ref{5c}) and (\ref{5d}). This has to be so, if one
wants to recover (\ref{5h}).

\section{Compatibility with known results}

We can follow \cite{zhang} and seek for a mean field solution in the presence
of a magnetic field $\Theta\left(  \mathbf{x}\right)  B=-\varepsilon
_{ij}\partial_{i}A_{j}$. Again the proposed form is:%
\begin{equation}
\phi\left(  x\right)  =\sqrt{\bar{\rho}},\qquad\mathbf{a}=-\mathbf{A},\qquad
a_{0}\left(  x\right)  =0,
\end{equation}
where $\bar{\rho}$ is the average particle density. The current takes the form
$j^{\mu}=\left(  \Theta\left(  \mathbf{x}\right)  \bar{\rho},\mathbf{0}%
\right)  $. The equations of motion in the CS sector are%
\begin{align}
&  \varepsilon_{ij}\partial_{0}a_{j}(x) =0 \; \;
\mbox{(identically satisfied)}\\
&  \varepsilon_{ij}\partial_{i}a_{j}(x) =\Theta\left(  \mathbf{x}\right)
B=\Theta\left(  \mathbf{x}\right)  \phi_{0}\frac{\theta}{\pi}\bar{\rho
}.\nonumber
\end{align}
Identifying the density of magnetic flux as $\rho_{{\Gamma}}=B/\phi_{0}$, one
obtains the condition for the validity of the mean field approximation%
\begin{equation}
\nu=\frac{\bar{\rho}}{\rho_{{\Gamma}}}=\frac{\pi}{\theta}=\frac{1}{2k+1}.
\end{equation}
This is exactly the same result obtained in \cite{zhang} for the filling factors.

Concerning the expression of the current in terms of the CS field, we can see,
using the CS equations of motion%
\begin{equation}
\frac{\delta S}{\delta a_{\mu}}=\frac{\delta}{\delta a_{\mu}}(S_{\mathrm{M}%
}+S_{\mathrm{CS}})=j_{\mathrm{M,}{\Gamma}}^{\mu}+\frac{\delta S_{\mathrm{CS}}%
}{\delta a_{\mu}}=0.
\end{equation}
So, $j_{\mathrm{M,}{\Gamma}}^{\mu}=$ $-\delta S_{\mathrm{CS}}/\delta a_{\mu}$.
Computing this last quantity, we obtain%

\begin{equation}
\frac{\delta S_{\mathrm{CS}}}{\delta a_{\mu}\left(  x\right)  }\equiv
ej_{\mathrm{CS}}^{\mu}=-\sigma_{xy}\varepsilon^{\mu\nu\rho}\partial_{\nu
}a_{\rho}\left(  x\right)  . \label{30}%
\end{equation}
This is precisely the final current obtained in \cite{maeda}, after the effect
of the chiral edge fermions has been taken into account. {In this paper, the
author integrates over chiral $(1+1)$-dimensional edge fermions to obtain an
effective action which, considered along with a bounded Chern-Simons action
(not gauge invariant), results in a gauge invariant theory. The anomalous term
in the current is cancelled and it remains only the contribution equal to that
of a non-bounded Chern-Simons action (which is gauge invariant). Our approach
leads directly to this result.}

{It must also be emphasised that the result (\ref{30}), contained in
\cite{wen,maeda}, means that the edge modes can not be identified with the
extra edge fermions. This is explicitly said by the author of ref.
\cite{maeda}, which follows the approach proposed in ref. \cite{wen}. Our
approach gives the same current without the need of introduction of any extra
degrees of freedom.}

All that remains is to see what happens at the edge using the equation
$\partial_{\mu}j_{\mathrm{M},{\Gamma}}^{\mu}=0$. This equation gives%
\begin{equation}
\partial_{\mu}\left(  \Theta\left(  \mathbf{x}\right)  j_{\mathrm{M}}^{\mu
}\right)  =\Theta\left(  \mathbf{x}\right)  \partial_{\mu}j_{\mathrm{M}}^{\mu
}+\left(  \partial_{\mu}\Theta\left(  \mathbf{x}\right)  \right)
j_{\mathrm{M}}^{\mu}=0. \label{separation}%
\end{equation}
Equation (\ref{separation}) implies two separated equations: one for the bulk
and other for the edge of the surface of the sample. We obtain,%
\begin{equation}
\partial_{\mu}j_{\mathrm{M}}^{\mu}=0,\; \mathrm{in\: the \: bulk,}
\label{bulk}%
\end{equation}
and%
\begin{equation}
j_{\mathrm{M},n}=n_{i}j_{\mathrm{M}}^{i}=0,\; \mathrm{in \: the \: edge,}
\label{edge}%
\end{equation}
where $n_{i}$ is a vector field that is normal to the boundary. Condition
(\ref{edge}) says that the current at the edge is only tangential, {and this
completes the identification of the edge modes relevant to the FQHE with the
matter current, as expected}.

\section{Conclusion}

Close inspection on the calculation done in references \cite{zhang,wen,maeda}
shows that one obtains, starting from $S$, given by equation (\ref{action})
and gauge invariant, the same value for the Hall conductance and edge current
that were obtained previously, starting with a theory which was not gauge
invariant and adding extra degrees of freedom in the form of chiral edge
fermions. So, there is no reason for the introduction of extra one-dimensional
chiral fermions circulating on the boundary.

\acknowledgments
G. L. S. Lima has been financially supported by CAPES (Brazil) during the
realization of this work.


\begin{thebibliography}{99}                                                                                               %


\bibitem {zhang}Zhang S. C., Hansson T. H. and Kivelson S., Phys. Rev. Lett.
\textbf{62 }(1989), {82}; Zhang S. C., Int. J. Mod. Phys. B \textbf{6} (1992),
{25}.

\bibitem {wen}Wen X. G., Phys. Rev. B \textbf{43} (1991), {11025}.

\bibitem {maeda}Maeda N., Phys. Lett. B \textbf{376} (1996), {142}.

\bibitem {girvin}Girvin S. M., \textit{The Quantum Hall Effect: Novel
Excitations and Broken Symmetries}, Les Houches Lecture Notes, Topological
Aspects of Low Dimensional Systems, Eds. Comtet A., Jolicoeur T., Ouvry S. and
David F., {Springer-Verlag, Berlin}, {2000}.

\bibitem {wang-goldman}Wang J. K. and Goldman V. J., Phys. Rev. Lett
\textbf{67} ({1991}), {749}.

\bibitem {yao}Yao W., Yang S. A. and Niu Q., Phys. Rev. Lett. \textbf{102}%
({2009}), {096801}.

\bibitem {hatsugai}Hatsugai Y., Sol. St. Comm. \textbf{149} ({2009}), {1061}.

\bibitem {boyarsky}Boyarsky A., Cheianov V. V. and Ruchayskiy O., Phys. Rev. B
\textbf{70} ({2004}), {235309}.

\bibitem {chang}Chang A. M., Rev. Mod. Phys.\textbf{75} ({2003}), {1449}.

\bibitem {basu}Basu B. and Bandyopadhyay P., Phys. Scr. \textbf{73} ({2006}),
{332}.
\end{thebibliography}
\end{document}